\documentclass[superscriptaddress,showpacs,preprintnumbers,amsmath,amssymb,twocolumn,prb]{revtex4}

\usepackage{epsfig}
\usepackage{dcolumn}
\usepackage{bm}
\usepackage{latexsym}
\usepackage{units}

\usepackage[]{natbib}
\usepackage[colorlinks=true, citecolor=blue]{hyperref}

\begin{document}
\preprint{APS/123-QED}

\title{Plasmonic crystals for ultrafast nanophotonics:\\ Optical switching of surface plasmon polaritons}
\author{M.~Pohl}
 \affiliation{Experimentelle Physik 2, Technische Universit\"at Dortmund, 44221 Dortmund, Germany}
\author{V.I.~Belotelov}
 \email{belotelov@physics.msu.ru}
 \affiliation{M.V. Lomonosov Moscow State University, 119991 Moscow, Russia}
 \affiliation{A.M. Prokhorov General Physics Institute, Russian Academy of Sciences, 119992 Moscow, Russia}
\author{I.A.~Akimov}
 \email{ilja.akimov@tu-dortmund.de}
 \affiliation{Experimentelle Physik 2, Technische Universit\"at Dortmund, 44221 Dortmund, Germany}
 \affiliation{A.F. Ioffe Physical-Technical Institute, Russian Academy of Sciences, 194021 St. Petersburg, Russia}
\author{S.~Kasture}
 \affiliation{Tata Institute of Fundamental Research, 400005, Mumbai, India}
\author{A.S.~Vengurlekar}
 \affiliation{Tata Institute of Fundamental Research, 400005, Mumbai, India}
\author{A. V. Gopal}
 \affiliation{Tata Institute of Fundamental Research, 400005, Mumbai, India}
\author{A.K. Zvezdin}
 \affiliation{A.M. Prokhorov General Physics Institute, Russian Academy of Sciences, 119992 Moscow, Russia}
\author{D.R. Yakovlev}
 \affiliation{Experimentelle Physik 2, Technische Universit\"at Dortmund, 44221 Dortmund, Germany}
 \affiliation{A.F. Ioffe Physical-Technical Institute, Russian Academy of Sciences, 194021 St. Petersburg, Russia}
\author{M. Bayer}
 \affiliation{Experimentelle Physik 2, Technische Universit\"at Dortmund, 44221 Dortmund, Germany}

\begin{abstract}
We demonstrate that the dispersion of surface plasmon polaritons in a periodically perforated gold film can be efficiently manipulated by femtosecond laser pulses with the wavelengths far from the intrinsic resonances of gold. Using a time- and frequency- resolved pump-probe technique we observe shifting of the plasmon polariton resonances with response times from 200 to 800~fs depending on the probe photon energy, through which we obtain comprehensive insight into the electron dynamics in gold. We show that Wood anomalies in the optical spectra provide pronounced resonances in differential transmission and reflection with magnitudes up to \unit[3]{\%} for moderate pump fluences of $\unit[0.5]{mJ/cm^2}$.
\end{abstract}





\maketitle

Nowadays plasmonics attracts much research interest in nanophotonics inspiring scientists to develop a new paradigm in data processing based on nanometallic circuitry \cite{Shalaev10,Brongersma10}. The key object of plasmonics is a surface plasmon polariton (SPP) - a coupled oscillation of the electromagnetic field and the electron plasma in metals \cite{Mai07}. Excitation of a SPP leads to significant electromagnetic energy localization near the metal-dielectric interface, thereby enhancing nonlinear effects and light-matter interaction. Current state-of-the-art in telecommunications requires plasmonics to be active, i.e. a possibility for control by means of an external stimulus must be provided on the order of a few nanoseconds or shorter \cite{Mac09}.

One of the approaches satisfying the strict criteria is magnetoplasmonics employing the influence of an external magnetic field on the SPP propagation constant \cite{Tem10,Bel11}. While the modulation efficiency in magnetoplasmonic structures may be as large as tens of percent the operation rate is limited by the magnetization dynamics occurring on the nanoseconds timescale. Transient changes in the optical properties of plasmonic structures can be also achieved via application of intense femtosecond laser pulses \cite{Mac09,Ext88,Sam09,Mus06,Cas10,Rot08,Rot09,Baida09,Dev08}. Here, the real and imaginary parts of the dielectric constant of a metal change in the matter of several hundreds of femtoseconds \cite{Fat98,Sun94,Fat00}. Since the propagation constant of SPPs is determined by the permittivity of metal and dielectric this opens new horizons for ultrafast control of SPPs. However, the high reflectivity of smooth gold surfaces prevents most of the incident electromagnetic energy from absorption in the metal. Here, SPPs can be exploited to provide electromagnetic energy localization near the metal-dielectric interface and consequently to increase the energy absorption in gold \cite{Gro95,Liu05}.

Recently the optical response of plasmonic crystals with periodically perforated gold or perforated dielectric on top of gold was investigated using a two-color pump-probe technique with the pump photon energy far from the SPP resonances \cite{Rot08,Rot09}. It was demonstrated that the differential reflectivity of the probe beam tuned in resonance with a SPP can be modulated up to \unit[10]{\%} on a sub-picosecond timescale. Large values of the modulation parameter were accomplished by using intense laser pulses with fluence $\Phi\sim $ \unit[50]{mJ/cm$^2$} and by tuning the resonances of the probed SPPs close to the $d$-band transitions in gold ($\hbar\omega_{ib} = 2.3$~eV \cite{Sun94}). However, the drawback of this approach is the large absorption in the vicinity of the $d$-band transition resulting in a huge intrinsic SPP damping. A similar obstacle, but with even larger optical losses, is present in Al plasmonic crystals \cite{Kir08}.

Here, we demonstrate that the dielectric constant of the metal in a plasmonic crystal can be efficiently modified at femtosecond time scale via SPP excitation, leading to ultrafast modification of the SPP dispersion. This is possible in the spectral range far from the intrinsic resonance frequencies of the metal system.  From the temporal and spectral dependence of differential reflectivity and transmission we extract comprehensive information on the electron energy relaxation in gold. We achieve \unit[3]{\%} probe intensity modulation with about an order of magnitude smaller pump fluence of $\Phi \approx 0.5$~mJ/cm$^2$ compared to previous works \cite{Rot09}.

\begin{figure}
\includegraphics[width=\linewidth]{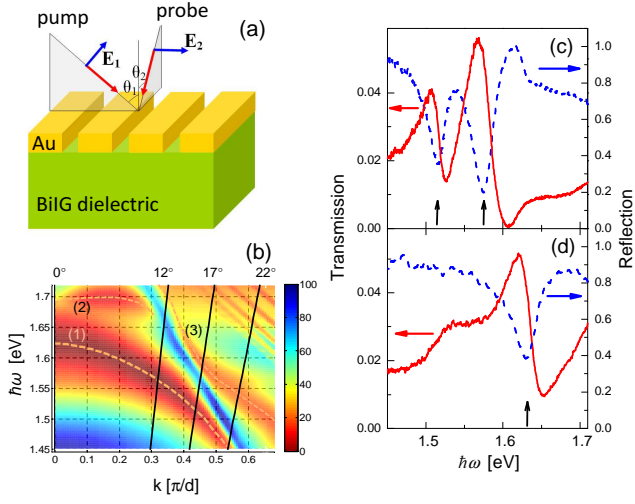}
\caption{(a) Scheme of the plasmonic crystal for active control of SPPs. (b) Dispersion of the SPP propagating at the BiIG/Au [(1),(2) dashed curves] and at the Au/air [(3) dashed curve] interfaces calculated by the S-matrix method. Black solid lines correspond to the dispersion of the plane waves in free space incident at $0^\circ$, $12^\circ$, $17^\circ$, and $22^\circ$ angle. (c,d) Experimentally measured transmission (solid lines) and reflection (dashed lines) spectra for the pump (c) and probe (d) configurations. Angle of incidence is $17^\circ$. Incident light is $p$-polarized in (c) and $s$-polarized in (d).}
\label{fig:Dispersion}
\end{figure}

The investigated plasmonic crystal is schematically shown in Fig.~\ref{fig:Dispersion}a. It comprises a periodically perforated, 120-nm-thick gold film deposited on top of a smooth ferromagnetic layer of rare-earth iron-garnet BiIG with \unit[2.5] ${\mu m}$ thickness. The period is $d=595$~nm with a groove depth of \unit[120]{nm} and an air groove width of \unit[110]{nm} [see \ref{A1}]. The time-resolved differential transmission and reflection were measured by a pump-probe technique using a Ti-Sa laser with a pulse repetition rate of about 80 MHz, center wavelength around 800~nm and a pulse duration of $\tau_D=\unit[30]{fs}$ [see \ref{A2}].  In the experiment the pump beam entered the sample in the plane perpendicular to the slits, while the probe beam was incident in the slit plane (Fig.~\ref{fig:Dispersion}a). The incidence angles $\theta_1$ and $\theta_2$ for both pump and probe were set to 17$^\circ$ with an aperture angle around $6^\circ$. The plasmonic crystal was designed to have SPP resonances in this configuration in the energy range \unit[1.5-1.7] {eV} for both the pump and probe beams.

The dispersion diagram for the SPP modes in this plasmonic crystal calculated with the scattering matrix (S-matrix) method \cite{Tik02, Bel10} reveals that the pump beam excites the SPPs in the 2-nd band of the Au/dielectric interface [curve (1) in Fig.~\ref{fig:Dispersion}b] and the 2-nd band SPP at the Au/air interface [curve (3) in Fig.~\ref{fig:Dispersion}b] with an energy around \unit[1.55]{eV}. On the other hand, the probe beam interacts with the SPPs in the 2-nd band of the Au/dielectric interface at \unit[1.63]{eV} around $k=0$. Zero order reflection and transmission spectra of the investigated structure measured using spectrally broad halogen lamp demonstrate Wood anomalies \cite{Wood} related to the SPPs. Fig.~\ref{fig:Dispersion}c,d show pronounced resonances with an asymmetric Fano-like shape \cite{Luk10}. The SPP energies can be determined from absorption spectra and are indicated by the arrows.

The dielectric constant of Au has contributions from conduction and bound electrons and can be written as
\begin{align}
	\label{eq:epsilon}
	\epsilon_m = 1-\frac{\omega^{2}_{p} }{\omega(\omega+i\gamma)}+\chi^{ib}_1+i\chi^{ib}_2,
\end{align}
where $\omega_{p}$ is the plasma frequency, $\gamma$ is the electron scattering rate, and $\chi^{ib}_1$ and $i\chi^{ib}_2$ are the real and imaginary parts of the interband susceptibility. While $\chi^{ib}_2$ is centered around the electron transition frequencies, $\chi^{ib}_1$ is spread in frequency and gives noticeable contributions to $\epsilon_m$ even far from the $d$-band transition in gold \cite{Pin04}. Optically induced ultrafast changes in $\gamma$, $\chi^{ib}_1$, and $i\chi^{ib}_2$ are consequences of modifications of the electron and lattice properties of gold, caused by electromagnetic energy absorption.

Intraband absorption gives rise to a strongly non-equilibrium electron distribution around the Fermi energy \cite{Sun94}. Due to electron-electron scattering, electron thermalization takes place leading to a hot Fermi distribution. Simultaneous to the internal thermalization electron-phonon interactions transfer the energy to the lattice in a matter of \unit[1] {ps}. This process leads to the transient changes in $\chi^{ib}_1$ . The values of $\Delta\chi^{ib}_1$ are approximately proportional to the integrated absorption of the laser pulse and evolves in parallel with the electron-lattice thermalization \cite{Gro95}. Intraband absorption also changes the scattering rate $\gamma$. However the temporal behavior of $\Delta \gamma$ is different from the one for $\Delta \chi^{ib}_1$. While $\Delta \chi^{ib}_1$ reaches its maximum as soon as the pump pulse is finished, $\Delta \gamma$ increases during the first hundred femtoseconds and then decreases on time scales much larger than 1~ps \cite{Fat00}. This is due to the fact that at the earlier stages $\Delta \gamma$ is mainly due to the heating of the electron and phonon systems, while for the subsequent stages it is dominated by energy relaxation of these systems via electron-phonon and phonon-phonon interactions \cite{Fat98}. Along with the excitation of the conduction electrons, there is also interband absorption as long as $\hbar \omega_{1} + \hbar \omega_{2} > \hbar \omega_{ib}$ which is satisfied for our case. It leads to ultrafast changes in $\chi^{ib}_2$ persisting shorter than a hundred femtoseconds \cite{Fat00}.

\begin{figure}
\includegraphics[width=7.5cm]{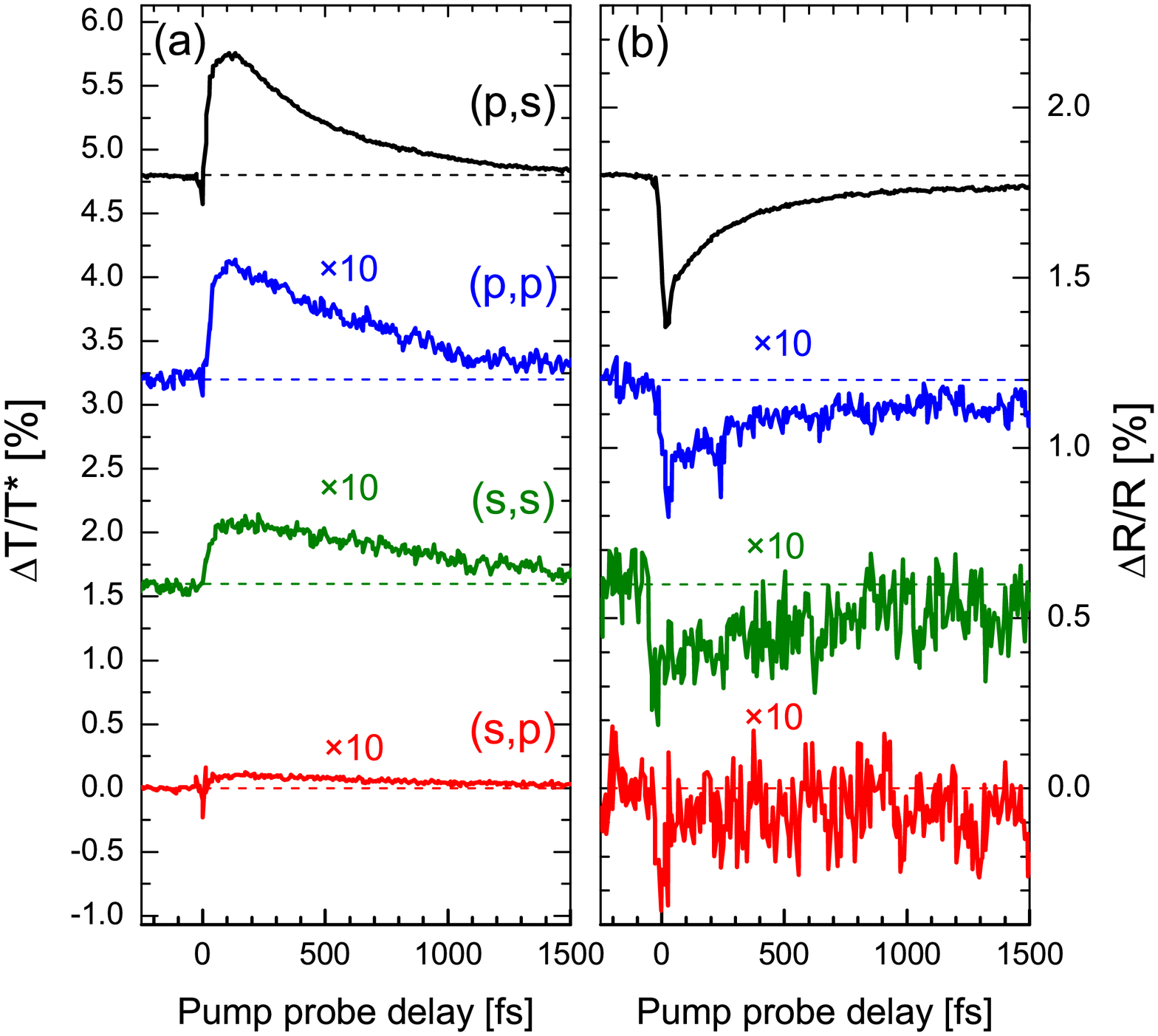}
\caption{Time-resolved differential transmission (a) and reflectivity (b) signals measured for different polarization configurations of pump and probe beams. The signals are shifted vertically for clarity (zero levels are shown by dashed lines). The signals for (p,p), (s,s), and (s,p) configurations are multiplied by 10. Pump fluence here and for the experimental results in all subsequent figures is \unit[0.5]{mJ/cm$^2$}.}
\label{fig:Polarization}
\end{figure}

Plasmonic crystals possess extraordinary optical transmission and, therefore, transient changes in the gold permittivity can be monitored not only in reflection but also in transmission. Since SPPs are TM-polarized they can be excited only by $p$-polarization of the pump beam or by $s$-polarization of the probe beam. The strongest transient signals of differential transmission $\Delta T/T$ and reflection $\Delta R/R$ are expected in the case when both pump and probe excite SPPs. Indeed, pump SPP allows significant electromagnetic energy absorption in gold, whereas probe SPP causes Fano resonances in $T$ and $R$ spectra. The latter increases the sensitivity of the transient signals due to the changes in the optical properties of gold. Experimental results prove this statement unambiguously (Fig.~\ref{fig:Polarization}). As expected, the signal is strongest for the fully plasmonic configuration $(p,s)$ ($p$- and $s$-polarizations of pump and probe, respectively). Compared to this configuration the signal strongly decreases for all other polarization combinations of pump and probe\footnote{Here, the $T$ of the probe for the non-plasmonic case is much smaller, and therefore for adequate comparison of all configurations, the transient signal in transmission was normalized by the transmission for the case of the $s$-polarized probe $T^*$.}.

The signal in the totally non-plasmonic case of $s$-polarized pump and $p$-polarized probe can hardly be seen. In the $(p,p)$ case the SPP is  excited by the pump beam but the changes in reflectivity and transmission are more than one order of magnitude smaller. This is explained by the weak sensitivity of both transmitted and reflected probe beam due to absence of resonances in the probed spectrum, and demonstrates the importance of SPP for transient optical property sensing. Finally, in the $(s,s)$ configuration a plasmonic resonance is present in the $T$ and $R$ spectra but the signal is still as small as in the $(p,p)$ configuration. This manifests the importance of SPP excitation by the pump beam, which provides an efficient trapping of the incident photons and their consecutive absorption in the gold.

The transients $\Delta T/T$ and $\Delta R/R$ versus time $t$ can be well described by a sum of two contributions
\begin{align}
	\label{eq:fit}
	\frac{\Delta I}{I}(t) = & D_{j} e^{ -\left[ \frac{t}{\sigma} \right] ^2 } +  \frac{A_{j}}{2} \left[ 1+ \text{erf} \left( \frac{t}{\sigma}-\frac{\sigma}{2\tau_{j}} \right) \right] e^{ \left[ \frac{\sigma}{2\tau_{j}} \right] ^2- \frac{t}{\tau_{j}}} + \nonumber \\
       & +  \frac{B_{j}}{2} \left[ 1+ \text{erf} \left( \frac{t}{\sigma} \right) \right]
\end{align}
where $I=R,T$, $j = R,T$, $\tau_j$ is the relaxation time, and $\sigma=\tau_D / \sqrt{2\ln 2}$. Here, the first term corresponds to a fast instantaneous component, which follows the autocorrelation function of the laser pulses. This component is also present in $\Delta R/R$ from flat gold when the pump and the probe are incident from the magnetic film side. Therefore, we attribute this signal to the non-linear optical response from the magnetic film related with optical transitions in octahedrally coordinated Fe$^{3+}$ ions \cite{Grishin}. The next contribution is described by the second and the third terms in Eq.~\ref{eq:fit}. It manifests a fast increase of the signal followed by a monotonous decay to a plateau. This contribution is associated with optically induced excitation of electrons and their energy relaxation in gold. The plateau is due to much slower relaxation processes in the phonon system. This contribution can be described by a multi-exponential decay convoluted with the apparatus function. For simplicity, we consider this optical response as a single exponential decay with amplitude $A_j$ and decay time $\tau_{j}$ reaching a plateau at the level $B_j$. In the remainder of this work we concentrate on transients related to the fast exponential decay.

\begin{figure}
\includegraphics[width=\linewidth]{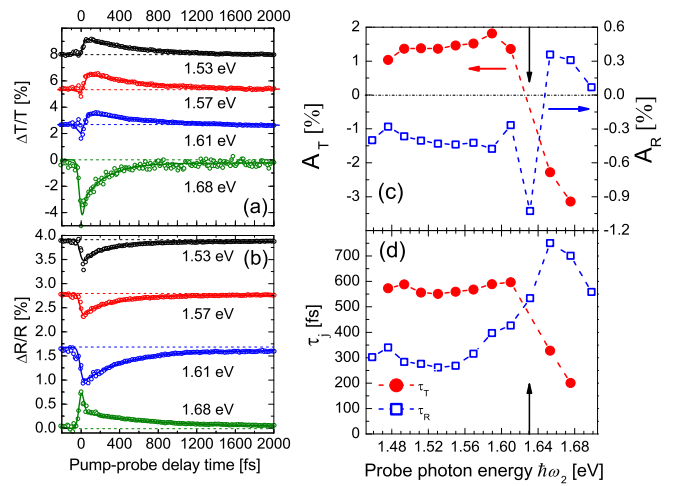}
\caption{Left panel: Transient transmission (a) and reflection (b) at different probe energies $\hbar\omega_2$=1.53~eV, 1.57~eV, 1.61~eV, 1.68~eV. The solid lines are fits with Eq.~\ref{eq:fit}. Right panel: Spectral dependencies of the amplitudes $A_j$ (c) and the decay times $\tau_j$ (d) of the transient signal in transmission ($j=T$, solid symbols) and reflection ($j=R$, open symbols) evaluated from fits of the experimental data. Vertical arrows indicate the energy of the probe SPP.}
\label{fig:TransientsFit}
\end{figure}

The polarization properties (see Fig.~\ref{fig:Polarization}) directly show the major role of SPPs for efficient energy absorption and high probe sensitivity. Consequently, the spectral dependence of the transient signals should also reflect resonant behavior and have extrema in the vicinity of the plasmonic resonances.  In order to check this, we introduced an electrically tunable interference filter (\unit[20]{meV} bandwidth) in the reflected/transmitted probe beam, directly in front of the photodetector. Thereby, we were able to probe the evolution of the SPP resonances after absorption of the pump pulse (see Fig.~\ref{fig:TransientsFit} a,b). Both $\Delta T/T$ and $\Delta R/R$ signals change sign around 1.60-\unit[1.66]{eV}. This energy region clearly corresponds to the SPP at the gold/BiIG interface probed in the experiment (see curve (1) in Fig.~\ref{fig:Dispersion}b).

In order to extract information about changes in $\gamma$, $\chi^{ib}_1$, and $\chi^{ib}_2$ we analyze the spectral dependence of the optical response using Eq.~\ref{eq:fit}. The spectral dependencies of the coefficients $A_{T}$ and $A_{R}$ show resonances with a sign change around the plasmonic resonance of the probe (Fig.~\ref{fig:TransientsFit} c). Such behavior can be explained by a shift of the Fano resonance along with a broadening, which indicates that both real and imaginary parts of $\epsilon_{m}$ change.  Surprisingly, the relaxation times $\tau_{T}$ and $\tau_{R}$  also vary with energy (Fig.~\ref{fig:TransientsFit} d). This indicates that the relative contributions of different parts of the transient permittivity in $\Delta T/T$ and $\Delta R/R$ are energy dependent.

\begin{figure}
\includegraphics[width=\linewidth]{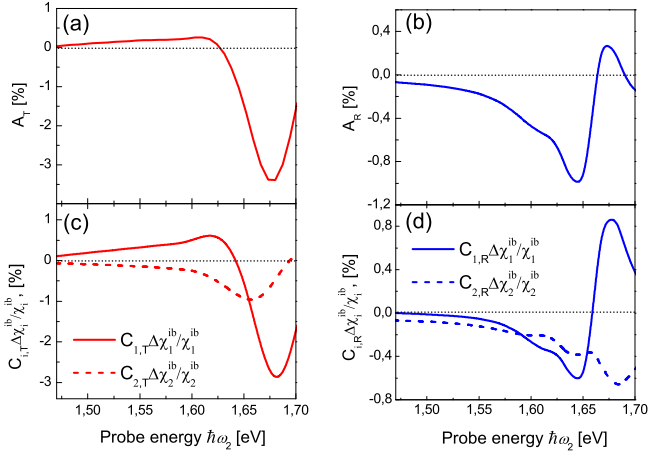}
\caption{Spectral dependencies of $A_T$ (a) and $A_R$ (b) obtained by RCWA modeling, to have the best correspondence with the experimental data. (c,d) Spectral dependencies of different contributions related with $\Delta \chi_1^{ib}$ ($i=1$, solid lines) and $\Delta \chi_2^{ib}$ ($i=2$, dashed lines) in accordance with Eq.~\ref{eq:combine}.}
\label{fig:Modeling}
\end{figure}

In the spectral region, which is addressed in experiment, $\Delta \gamma$ influences only the imaginary part of $\epsilon_{m}$. As a result, $\Delta \gamma$ and $\Delta \chi_2^{ib}$ should have similar spectral dependencies. However, as discussed above $\Delta \chi_2^{ib}$ relaxes much faster than $\Delta \gamma$ does. Since $\Delta \gamma$ relaxes during times much longer than several picoseconds (which is out of the observation range) it causes a plateau in the transient signals described by $B_j$ in Eq.~\ref{eq:fit}. On the contrary, $\Delta \chi_2^{ib}$ relaxes much faster and contributes to the exponential decay with the amplitudes $A_j$ in Eq.~\ref{eq:fit}. Similarly, $\Delta \chi_1^{ib}$ also contributes to $A_j$. Thus, in the perturbation regime the $A_j$ can be written as linear combination of $\Delta \chi_1^{ib}$ and $\Delta \chi_2^{ib}$ with frequency dependent coefficients $C_{1,j}(\omega )$ and $C_{2,j}(\omega )$:

\begin{align}
\label{eq:combine}
A_{j}(\omega ) = C_{1,j}(\omega ) \frac{\Delta \chi_1^{ib}}{\chi_1^{ib}} + C_{2,j}(\omega ) \frac{\Delta \chi_2^{ib}}{\chi_2^{ib}} .
\end{align}
In our experiments the pump energy is far from the $d$-band transitions in gold so ${\chi_1^{ib}}$ and ${\chi_2^{ib}}$ can be considered constant for the energy range of interest. Spectral behavior of the coefficients in eq~\ref{eq:combine} can be determined by modeling the $T$ and $R$ spectra applying a rigorous coupled waves analysis (RCWA) \cite{Moh95}. Equation \ref{eq:combine} gives the best fits of $A_{j}(\omega)$ for ${\Delta \chi_1^{ib}}/{\chi_1^{ib}} = 0.4\times 10^{-2}$, and ${\Delta \chi_2^{ib}}/{\chi_2^{ib}} = 2.6\times 10^{-2}$ (Fig.~\ref{fig:Modeling}).

It is seen from Fig.~\ref{fig:Modeling} that the relative contributions of  $\Delta \chi_1^{ib}$ and  $\Delta \chi_2^{ib}$  determined by $C_{1,j}(\omega ) {\Delta \chi_1^{ib}}/{\chi_1^{ib}}$ and $C_{2,j}(\omega ) {\Delta \chi_2^{ib}}/{\chi_2^{ib}}$ are energy dependent. In the energy range of 1.45 - \unit[1.60]{eV}, the term $C_{1,T}(\omega ) {\Delta \chi_1^{ib}}/{\chi_1^{ib}}$ prevails in absolute value over the term $C_{2,T}(\omega ) {\Delta \chi_2^{ib}}/{\chi_2^{ib}}$ and the term $C_{2,R}(\omega ) {\Delta \chi_2^{ib}}/{\chi_2^{ib}}$ prevails over the term $C_{1,R}(\omega ) {\Delta \chi_1^{ib}}/{\chi_1^{ib}}$. At higher energies of 1.60 - \unit[1.66]{eV} these relations get reversed. Since different contributions have different relaxation times this leads to the energy dependence of $\tau_j$. As discussed above, $\Delta \chi_2^{ib}$ relaxes in less than a hundred femtoseconds, which diminishes $\tau_j$. This explains the experimentally observed drop of $\tau_{T}$ for larger energies where $\Delta \chi_2^{ib}$ contributes in Eq.~\ref{eq:combine} stronger (see Fig.~\ref{fig:TransientsFit} (d)). The decrease of $\tau_{R}$ for smaller energies can be explained similarly by stronger contribution of $\Delta \chi_2^{ib}$.

Furthermore, the significantly different temporal behaviors of $\Delta \gamma$ and $\Delta \chi_2^{ib}$ allow us even to discriminate their input in the differential signals in spite of the similarity of their spectral contributions. The coefficients $B_j$ in Eq.~\ref{eq:fit} give a ${\Delta \gamma}/{\gamma} = 3.0\times 10^{-2}$ corresponding to  ${\Delta \gamma}/{\Delta\chi_2^{ib}} = 1.15$. This nicely agrees with the data in \cite{Fat00}, providing ${\Delta \gamma}/{\Delta\chi_2^{ib}} \approx1$.

In conclusion, we show that the optical properties of plasmonic crystals can be tuned by femtosecond laser pulses. This opens a nice possibility for the control of the SPP dispersion in the plasmonic crystal. As the result SPP propagation through the plasmonic crystal is modified and corresponding differential transmission and reflection of the order of \unit[3]{\%} is observed. Moreover, the response time of plasmonic crystals can be varied widely from 200 to 800~fs by tuning the relative spectral positions of the probe and the SPP resonance. The results are important for operation both in near and far optical fields. In the near field it provides the possibility for ultrafast all-optical control of SPPs propagating through a plasmonic crystal. In the far field it allows one to modify the reflection and transmission spectra of the plasmonic crystal, thus making this material of great importance for nanophotonic applications. Moreover, the plasmonic crystal is also interesting for fundamental research, as it provides Wood's anomalies which allow clear distinctions among the contributions of the different gold permittivity components. An analysis of transient optical spectra along with a rigorous solution of Maxwell's equations provide sufficient information for determining magnitude and dynamics of $\Delta \chi_1$, $\Delta \chi_2$, and $\Delta \gamma$.

The authors are grateful to M.~Betz for useful discussions. The work is partly supported by the Deutsche Forschungsgemeinschaft,
Russian Foundation of Basic Research (RFBR), DST (India), the
Russian Presidential Grant MK-3123.2011.2, and by the Federal
Targeted Program (No. 16.740.11.0577).

\appendix

\section{Sample preparation}\label{A1} The magnetic part of the magnetoplasmonic structure is a 2.5 $\mu$m thick bismuth-substituted rare-earth iron garnet film of composition Bi$_{0.4}$(YGdSmCa)$_{2.6}$(FeGeSi)$_{5}$O$_{12}$ grown by liquid phase epitaxy from Bi$_{2}$O$_{3}$ : PbO : B$_{2}$O$_{3}$ melt on the Gd$_{3}$Ga$_{5}$O$_{12}$ substrate with orientation (111). The film possesses a uniaxial magnetic anisotropy in the direction perpendicular to the film plane. The specific Faraday rotation is \unit[0.46]{deg/$\mu$m} at a wavelength of \unit[633]{nm}. The magnetoplasmonic sample of structure shown in Figure 1a was fabricated by the thermal deposition of the gold layer on the bismuth-substituted rare-earth iron garnet film and subsequent electron beam lithography combined with the reactive ion etching in Ar plasma. The sample was characterized by AFM and SEM imaging. The following grating parameters were obtained: Gold layer height $h= \unit[120]{nm}$, period $d= \unit[595]{nm}$ and the air groove width $r= \unit[110]{nm}$. \\

\section{Experimental technique}\label{A2}
A Ti-Sapphire self-mode locked laser with a pulse repetition rate of about 80 MHz was used as the source of ultra-short optical pulses with the center wavelength around 800~nm and a bandwidth of 80~nm. The laser beam was sent through a pulse shaper and a compressor, allowing for an overall time resolution of about \unit[40]{fs}. Subsequently, the laser beam was split into the pump and probe beams. The pump beam passed a mechanical delay line and was modulated with a mechanical chopper at a frequency of 1-5 kHz. Both pump and probe beams were focused onto the plasmonic crystal with a reflective microscope objective into the spot of about 5 $\mu$m diameter. The intensities of pump and probe pulses were set to 500 and 50 $\mu$J/cm$^2$, respectively. The transmitted or reflected probe beam was homodyne detected with a balanced photodiode and a lock-in amplifier. \\

\bibliography{references}

\end{document}